\documentclass[fleqn,10pt]{wlscirep}
\usepackage[utf8]{inputenc}
\usepackage[T1]{fontenc}
\usepackage{lineno}
\usepackage{siunitx}
\usepackage{graphicx}
\usepackage{nameref}
\usepackage{hyperref}
\usepackage{algorithm}
\usepackage{algorithmic}
\usepackage[capitalize]{cleveref}
\usepackage{amsmath}
\usepackage{bm}
\usepackage{enumitem}
\usepackage{lipsum}

\title%[EBC PLIF]
{
Odour sensing in turbulent plumes with high-speed electronic nose and non-invasive ground truth
}

\author[1,2,3,$\dagger$]{Nik Dennler}
\author[4]{Elle Stark}
\author[1]{Saimon Collaku}
\author[4]{Lars Larson}
\author[2,5]{Andr{\'e} van Schaik}
\author[3]{Michael Schmuker}
\author[4]{John Crimaldi}
\author[1]{Andreas T. G{\"u}ntner}
\author[4, $\dagger$]{Aaron True}

\affil[1]{\footnotesize Human-centered Sensing Laboratory, Department of Mechanical and Process Engineering, ETH Zurich, Zurich, Switzerland}
\affil[2]{\footnotesize International Centre for Neuromorphic Systems, Western Sydney University, Kingswood, NSW, Australia}
\affil[3]{\footnotesize Biocomputation Group, Centre for Data Innovation Research, University of Hertfordshire, Hatfield, United Kingdom}
\affil[4]{\footnotesize Department of Civil, Environmental and Architectural Engineering, University of Colorado Boulder, Boulder CO, USA}
\affil[5]{\footnotesize International Centre for Neuromorphic Systems, University of Manchester, Manchester, United Kingdom}
\affil[$\dagger$]{\footnotesize Correspondence: dennlern@ethz.ch, aaron.true@colorado.edu}

\begin{abstract}
Chemical sensing in real-world environments requires resolving rapidly fluctuating and spatially heterogeneous concentration fields. 
However, these dynamics are strongly distorted by widely used, low-cost metal-oxide (MOx) gas sensors, whose thermal and surface-kinetic response acts as a low-pass filter on the underlying concentration signal.
Quantifying and compensating for these effects remains challenging, largely due to the lack of benchmark datasets that simultaneously capture the spatiotemporal structure of turbulent odour fields and the time-resolved response of point sensors.
Here, we present a dataset combining planar laser-induced fluorescence (PLIF) measurements of an acetone tracer plume with synchronised recordings from a custom, kilohertz-rate microelectromechanical (MEMS) MOx electronic nose deployed in a laboratory wind tunnel. The PLIF system provides quantitative, two-dimensional concentration fields at high spatial and temporal resolution, while the co-located e-nose records film resistance, heater currents, and environmental parameters with aligned timestamps. The dataset enables quantitative assessment of sensor dynamics, development and benchmarking of reconstruction and deconvolution algorithms, and data-driven modelling of plume structure. All recordings, metadata, calibration files, and example analysis scripts are released in open, platform-independent formats. Together, these provide a valuable reference for researchers working in odour-guided robotics, environmental monitoring, computational fluid dynamics, and neuromorphic sensing, supporting the design and evaluation of high-speed odour-sensing systems.
\end{abstract}
\begin{document}

\flushbottom
\maketitle

\thispagestyle{empty}

\section*{Background and Summary}\label{sec:background}
%
% Artificial olfaction has advanced rapidly in recent years, with electronic noses (e-noses) emerging as promising tools for odour-guided robotics \cite{ishida_chemical_2012} and automated environmental monitoring \cite{yamazoe_development_1995}. 
Artificial olfaction has advanced rapidly in recent years, with electronic noses (e-noses) emerging as compact, low-cost, and scalable sensing platforms suitable for widespread deployment in odour-guided robotics \cite{ishida_chemical_2012} and automated environmental monitoring \cite{yamazoe_development_1995}.
Yet, successfully operating in real-world environments requires adequate resolution of the complex dynamics of turbulent odour plumes. 
Unlike controlled laboratory headspace samples, odour plumes are highly intermittent, filamentous, and characterised by rapid concentration fluctuations at frequencies that span tens to hundreds of hertz\cite{celani_odor_2014}. 
Sampling these at sufficiently high temporal resolution reveals informative odour features \cite{pannunzi_odor_2019}, such as the amplitude, intermittency or encounter frequency of odour packets, or the degree of temporal correlation between two odour signals. 
These features encode ecologically and operationally relevant information\cite{crimaldi_active_2022, ouyang_simple_2024}, including plume dimensions \cite{fackrell_effects_1982}, distance to source\cite{mafra-neto_fine-scale_1994, riffell_flower_2014, schmuker_exploiting_2016}, and inter-source-separation distances\cite{hopfield_olfactory_1991, ackels_fast_2021, tootoonian_quantifying_2025}. % and other source localisation cues. 
However, the metal-oxide (MOx) gas sensors most commonly used in e-noses act as strong physical low-pass filters: their thermal and surface-kinetic dynamics attenuate fast fluctuations\cite{korotcenkov_role_2008}, making it challenging to infer the underlying gas concentration time series. 
Microelectromechanical (MEMS)-fabricated MOx sensors may partially mitigate this limitation by reducing the thermal mass of heater substrate and sensor film \cite{gardner_micromachined_2023}, where recent work on tight control of the sensor operation temperature demonstrates the feasibility of high-speed sensing \cite{dennler_high-speed_2024}. 
%on tightly temperature-controlled MOx sensor arrays has shown that high-speed sensing is feasible with current gas sensors\cite{dennler_high-speed_2024}. 
Yet, quantitative validation of how closely these sensors may track true plume dynamics is limited, % by the absence of appropriate ground truth.
as the reconstruction of time-resolved gas concentrations from the low-pass-filtered signals of MEMS MOx sensors remains an open algorithmic challenge. 

Progress in this area is hampered mainly by the lack of benchmark sensor datasets\cite{france2025position}, particularly those capable of reconciling the low-pass sensor response with the true spatiotemporal structure of turbulent odour plumes. %, requiring a noninvasive ground-truth of adequate spatial and temporal resolution.
%
%Previous gas sensor studies have primarily focused on %collecting and 
%curating datasets %under controlled conditions 
%for benchmarking gas discrimination algorithms \cite{Vergara2013, fonollosa_chemical_2014, robin_deep_2022, dennler_high-speed_2024} or for quantifying sensor drift \cite{vergara_chemical_2012, kumar_gas_2023, worner_long-term_2025}. Typically, multiple analytes at various concentrations are tested in reproducible laboratory settings.
%In these efforts, typically been on testing multiple analytes at various concentrations in reproducible laboratory settings. 
%
%Consequently, most available datasets (apart from a single turbulent wind-tunnel dataset \cite{Vergara2013}, itself subject to notable limitations \cite{dennler_drift_2022, dennler_limitations_2024}) use highly controllable yet artificial odour stimuli, i.e., odour pulses with abrupt on and off behaviour. 
%
Previous gas sensor studies have primarily focused on curating datasets for benchmarking gas discrimination algorithms \cite{Vergara2013, fonollosa_chemical_2014, robin_deep_2022, dennler_high-speed_2024} or quantifying sensor drift \cite{vergara_chemical_2012, kumar_gas_2023, worner_long-term_2025}, typically under reproducible laboratory conditions. 
Consequently, most available datasets---apart from a single turbulent wind-tunnel dataset \cite{Vergara2013}, itself subject to notable limitations \cite{dennler_drift_2022, dennler_limitations_2024}---rely on highly controllable yet artificial odour stimuli, such as pulses with abrupt on--off behaviour.
However, turbulent scalar transport produces varied concentration gradients, strong intermittency, and heavy-tailed statistics\cite{dennler_neuromorphic_2025}, all of which vary with flow conditions and source configuration. 
Further, these datasets offer no dedicated concentration ground truth from which the sensors' filtering characteristics could be quantified. % quantitatively inferred.

Experimental measurements of turbulent gas plumes typically rely either on single-point sensors that provide limited spatial context \cite{justus_measurement_2002, drewnick_design_2012}, on optical methods without co-registered sensor measurements \cite{hargather2010natural, connor_quantification_2018, peng_visualization_2018}, or on combinations of optical setups and photo-ionisation detectors (PIDs) \cite{alvarez-salvado_elementary_2018}, where the latter are known to actively distort the flow field and embedded odour structure\cite{true_distortion_2022}. %due to their active sampling principle \cite{true2022distortion}.
%
%Meanwhile, computational fluid dynamics and scalar dispersion models may reproduce certain features of turbulent odour plumes\cite{farrell_filament-based_2002, monroy_gaden_2017, singh_emergent_2023}, 
%but do so at the cost of significant computational resources. To reduce these costs, embedded turbulence models are typically used to approximate physical processes at the smallest spatial scales and fastest time scales of the flow and odour fields \cite{burton2008nonlinear, donzis2010resolution}. Thus, they struggle 
%however, they struggle 
%to generate fully realistic fluctuations at the spatial and temporal resolutions required for adequately validating algorithms on odour signal reconstruction or olfactory navigation. %, or the training of data-driven odour sensing models. 
Meanwhile, computational fluid dynamics and scalar dispersion models may reproduce certain features of turbulent odour plumes \cite{farrell_filament-based_2002, monroy_gaden_2017, singh_emergent_2023}, but resolving the full range of spatial and temporal scales relevant to odour fluctuations is computationally prohibitive. Consequently, practical simulations rely on embedded turbulence models to approximate small-scale and fast dynamics \cite{burton2008nonlinear, donzis2010resolution}, limiting their ability to reproduce realistic intermittency and concentration statistics. This constrains their usefulness for the quantitative validation of odour signal reconstruction and olfactory navigation algorithms.
Thus, %despite extensive literature on turbulent dispersion, 
there remains no openly available dataset that jointly and non-invasively captures ground-truth concentration fields and fast gas sensor responses under realistic turbulent conditions.

\begin{figure*}[t]
    \centering
    \includegraphics[width=0.99\textwidth]{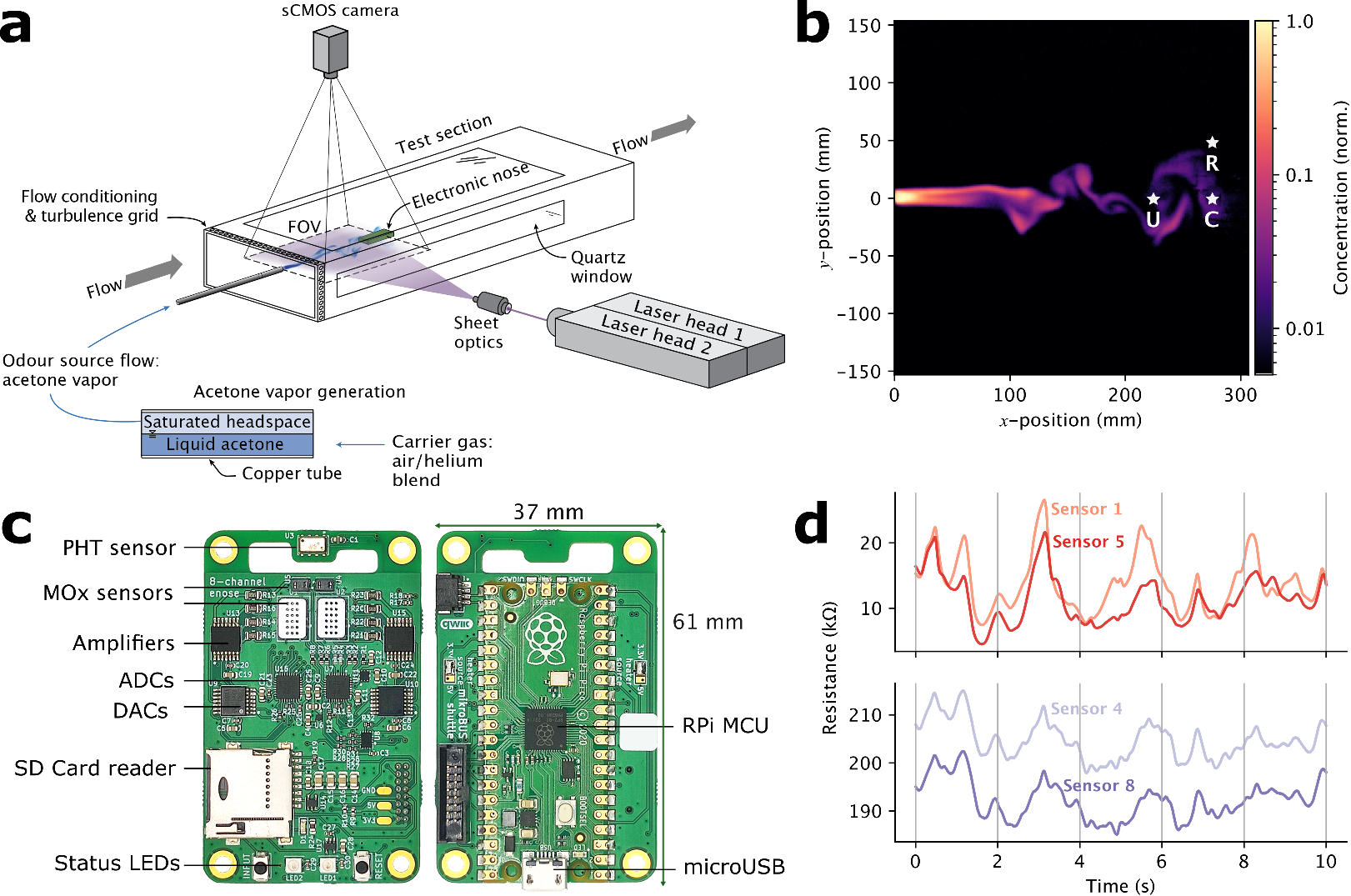}
    \caption{
    Experimental setup for odour measurements in turbulent flow.
    \textbf{a} Simultaneous monitoring of acetone in a wind tunnel using planar laser-induced fluorescence (PLIF) and a high-speed electronic nose (e-nose). 
    \textbf{b} PLIF output (source-normalised concentration), including e-nose locations: centre (\texttt{C}), \SI{5}{\centi\metre} upstream (\texttt{U}), and \SI{5}{\centi\metre} right (\texttt{R}).
    \textbf{c} E-nose printed circuit board (PCB), including pressure-humidity-temperature (PHT) sensor, metal-oxide (MOx) gas sensors, voltage amplifiers, analog-to-digital converters (ADCs), digital-to-analog converters (DACs), SD card reader, status LEDs, microUSB port and a Raspberry Pi (RPi) microcontroller (MCU).
    % \textbf{d} PLIF time series data extracted from sensor location, illustrating "bouts" and "blanks". 
    \textbf{d} Raw e-nose data snippet, for four MOx sensors.
    }
    \label{fig:ex1}
\end{figure*}

To address this gap, we developed a combined planar laser-induced fluorescence (PLIF) and high-speed e-nose measurement system in a turbulent wind-tunnel environment. 
The PLIF system (\cref{fig:ex1}a) provides quantitative, two-dimensional tracer concentration fields (here: acetone vapour) at high spatial and temporal resolution (\cref{fig:ex1}b), capturing the fine-scale filamentary structures characteristic of turbulent plumes \cite{connor_quantification_2018}. Co-located with the PLIF imaging plane, a custom eight-channel MOx array (\cref{fig:ex1}c) \cite{dennler_high-speed_2024} records local gas-sensor conductance at kilohertz rates (\cref{fig:ex1}d), together with temperature, humidity, and pressure. 
Hardware-level synchronisation between laser, camera, and e-nose, as well as spatio-temporal alignment between the resulting data modalities, enable direct comparison between the gas sensor responses and the respective tracer concentration. 
%, supporting quantitative assessment of sensor dynamics and evaluation of reconstruction or deconvolution approaches for recovering the original odour signal. 
%

Here, we describe a curated dataset comprising time-resolved PLIF concentration fields, co-registered high-frequency e-nose signals, and all associated metadata and calibration information. These resources enable direct, time-resolved quantification of gas sensor dynamics and allow falsifiable evaluations of reconstruction and deconvolution approaches for recovering the underlying odour signal from low-pass filtered measurements. 
Crucially, the dataset allows for empirically characterising the transfer function\cite{monroy_overcoming_2012, martinez_fast_2019, drix_resolving_2021} between turbulent concentration fluctuations and realistic MOx sensor responses under controlled yet representative flow conditions. %, thereby revealing both the information preserved by the sensors and the limits of what can be reconstructed. 
Further, it enables the development and benchmarking of sensor models, signal-processing pipelines, and machine-learning methods for plume inference under realistic turbulent statistics. %, rather than idealised pulse-based stimuli, as well as data-centric studies of turbulent dispersion.
By linking dense and non-invasive ground-truth measurements with co-located high-speed gas sensor recordings, the dataset constitutes a reusable experimental benchmark for olfactory navigation and robotics \cite{marjovi_multi-robot_2011, vincent_investigation_2019, france_olfactory_2025}, environmental monitoring \cite{collier2019understanding, khorramifar2023environmental}, and neuromorphic olfaction \cite{diamond_classifying_2016, jurgensen_neuromorphic_2021, han2022artificial, dennler_neuromorphic_2025}. 
Beyond these areas, the dataset is relevant to studies of turbulent scalar transport \cite{shraiman_scalar_2000}, inverse problems and system identification \cite{vogel02, ljung_system_1998}, chemical sensor design \cite{gardner_micromachined_2023}, as well as multimodal and physics-informed machine learning \cite{karpatne_theory-guided_2017}. % where paired ground-truth concentration fields and realistic sensor responses enable quantitative validation that is otherwise difficult to achieve.
%
%More broadly, it provides shared experimental infrastructure for reproducible, quantitatively grounded research on high-speed chemical sensing in turbulent environments.
%
%quantitative assessment of sensor dynamics and evaluation of reconstruction or deconvolution approaches for recovering the original odour signal. They enable the development and benchmarking of sensor models, signal-processing pipelines, and machine-learning methods for plume inference, as well as data-centric studies of turbulent dispersion. By linking dense optical measurements with realistic MOx responses under controlled yet representative turbulent flow conditions, the dataset provides a reusable benchmark for olfactory robotics \cite{ishida_chemical_2012, marjovi_multi-robot_2011, vincent_investigation_2019, france_olfactory_2025}, environmental monitoring \cite{yamazoe_development_1995, collier2019understanding, khorramifar2023environmental}, and neuromorphic olfaction\cite{diamond_classifying_2016, jurgensen_neuromorphic_2021, han2022artificial, dennler_neuromorphic_2025}, and facilitates reproducible research in high-speed chemical sensing.

% \newpage
\section*{Methods}
\subsection*{Turbulent Flow Facility}
A custom-made high-speed e-nose and a PLIF setup were employed to quantify the instantaneous tracer-gas concentrations in a low-speed wind tunnel. 
The flow facility consists of a low-speed recirculating wind tunnel with length 2.8 m, width 1.4 m, and depth 0.25 m. Airflow is driven by a variable-speed axial fan with flow conditioning using a honeycomb panel. 
%The flow facility is based on a previous PLIF design \cite{connor_quantification_2018}, but modified to a recirculating configuration. % and fitted with a fractal turbulence grid for enhanced turbulence intensity. 
%
The rectangular test section measures \SI{50}{\centi\metre}~(width)~$\times$~\SI{25}{\centi\metre}~(height)~$\times$~\SI{100}{\centi\metre}~(streamwise), and features an optically clear, \SI{6}{\milli\metre}-thick borosilicate top for \emph{en face} imaging. The laser sheet enters through a \SI{1.8}{\centi\metre}~$\times$~\SI{30}{\centi\metre} slit on the test section side wall, covered by a \SI{3}{\milli\metre}-thick quartz pane that allows transmission of the 266 nm light. 
Turbulent fluctuations were generated using a fabricated aluminium square-pattern fractal turbulence grid with blockage ratio $\sigma \approx 0.5$ and mesh Reynolds numbers $\mathrm{Re_M}$ of approximately 200, 300, and 400 for average wind speeds of 10, 15, and 20 cm/s, respectively. Additional grid design details include three fractal iterations (\(N=3\)), with bar lengths \(L_1=16\)~cm, \(L_2=8\)~cm, and \(L_3=4\)~cm, producing turbulence that decays downstream of the grid. Owing to the rectangular wind-tunnel cross-section, the number of repeated patterns was defined over the central \(25\times25\)~cm region of the grid (\(B=4\)), yielding a space-filling configuration with fractal dimension \(D_f=\log B/\log(1/R_L)=2\) for a bar-length ratio \(R_L=1/2\). The effective mesh size and thickness ratio, which govern the resulting flow properties \cite{hurst_scalings_2007}, were \(M_{\mathrm{eff}}=3\)~cm %=(4T^2/P)\sqrt{1-\sigma}=3\)~cm (with \(T^2=1250~\mathrm{cm^2}\), \(P=1175\)~cm, and \(\sigma\approx0.5\)) 
and \(t_r=2.9\)%t_{\max}/t_{\min}=2.9\) %(with \(t_{\max}=1.75\)~cm and \(t_{\min}=0.6\)~cm)
, respectively.

\subsection*{Planar laser-induced fluorescence setup}
% Flow facility
% A planar laser-induced fluorescence (PLIF) was employed to quantify the instantaneous tracer-gas concentrations in a benchtop-scale, low-speed wind tunnel. 
% %
% The flow facility is based on a previous design \cite{connor_quantification_2018}, but modified to a recirculating configuration and fitted with a fractal turbulence grid for enhanced isotropy. 
% %
% The rectangular test section measures \SI{50}{\centi\metre}~(width)~$\times$~\SI{25}{\centi\metre}~(height)~$\times$~\SI{100}{\centi\metre}~(streamwise), and features a optically clear, \SI{6}{\milli\metre}-thick borosilicate top for \emph{en face} imaging.

% Stimulus / Fluorescent tracer
As a fluorescent tracer, acetone vapour was selected for its high vapour pressure, low toxicity, and favourable photophysical properties (i.e. strong UV absorption at \SI{266}{\nano\metre} and broad fluorescence band), with fluorescence intensity exhibiting a near-linear response to concentration over the operating range \cite{lozano_acetone_1992}. To generate the acetone vapor, we bubbled a blended carrier gas through a long copper tube partially filled with liquid acetone (\textit{Fisher Chemical}, 99.5\% purity). The long residence time of the tube ensures saturation of the carrier gas, which consists of blended air and helium (\textit{AirGas USA}, 99.995\% purity) to produce a neutrally buoyant tracer plume. Precise volumetric flow rates of air and helium for each experiment are logged in the project HDF5 metadata.
Based on the laboratory conditions (temperature of 19-\SI{20}{\celsius}, atmospheric pressure of \SI{840}{\hecto\pascal}) and vapourisation efficiencies in the range 95--99\%,  the acetone source concentration was estimated as \(9.6\ \pm 0.4~\mathrm{mol\,m^{-3}}\).
%

%
% Laser excitation & imaging
Excitation was provided by two alternating Nd:YAG \SI{1064}{\nano\metre} laser heads, fitted with fourth-harmonic generator crystals to produce \SI{266}{\nano\metre}, \SI{5}{\nano\second} pulses at a frequency of \SI{20}{\hertz} (each head operated at \SI{10}{\hertz}). 
The combined beams were shaped into a planar light sheet (thickness $\approx$ \SI{500}{\micro\metre}) by a set of cylindrical and spherical optics, and aligned parallel to the test-section floor, bisecting the acetone injection port. 
Fluorescence was recorded from above using a Zyla~4.2+ sCMOS camera (monochrome, \SI{16}{\bit}, \num{2048} px$\times$\num{2048} px native resolution; \SI{2}{\times} hardware binning to yield \num{1024} px$\times$\num{1024} px images), selected for its low readout noise, high quantum efficiency, and high dynamic range.
Acquisition timing (i.e. laser lamp firing, camera exposure, and trigger signals for temporal synchronisation) was managed by custom LabVIEW routines. %, ensuring repeatable interleaving of PLIF frames and e-nose measurements. 

\subsection*{High-speed electronic nose}
\begin{figure*}[t]
    \centering
    \includegraphics[width=0.99\textwidth]{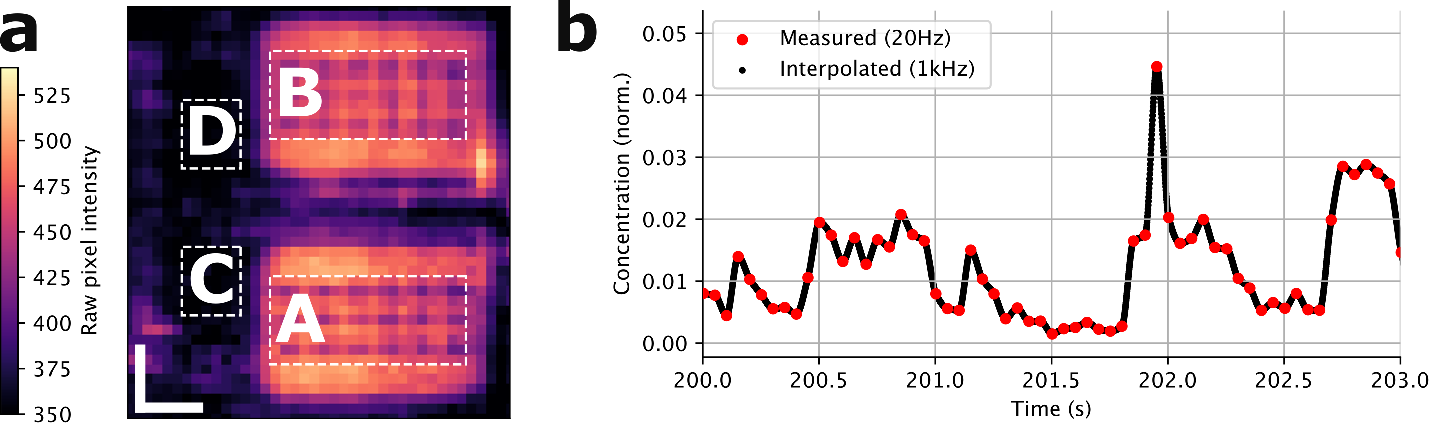}
    \caption{
    Data post-processing methods. 
    %\textbf{a} Frame number vs. raw intensity near the acetone source, illustrating how the signal requires about $\approx 400$ frames to stabilise.
    %\textbf{b} Linear regression of PLIF source-normalised intensity at a sensor interior (IR) location vs. an adjacent exterior (ER) area. The fits' slope of 0.85 quantifies the reflectance property of the \textit{SGX} \textit{MiCS-6814} polished aluminium housing.
   \textbf{a} Spatial alignment of six \textit{SGX} \textit{MiCS-6814} elements in two housings (sensor 1,2,3 in zone A and sensor 5,6,7 in zone B) and two \textit{ScioSense} \textit{CCS801} units (sensor 4 in zone C and sensor 8 in zone D), based on acquired flat-field image. Scale bar shows \SI{2}{\milli\metre} x \SI{2}{\milli\metre}. 
    \textbf{b} Area-averaged PLIF data, oversampled via piecewise cubic Hermite interpolating polynomial (PCHIP) \cite{fritsch1980monotone}.
    }
    \label{fig:processing}
\end{figure*}

The high-speed e-nose used in this experiment is described elsewhere \cite{dennler_high-speed_2024}, and centred on a Raspberry Pi Pico microcontroller, which handles real-time control, data acquisition and on-board storage to a microSD card. Its sensing front-end is comprised of four sensor housings with a total of eight MEMS metal-oxide (MOx) gas sensors---six \textit{SGX} \textit{MiCS-6814} elements in two housings (\cref{fig:processing}c, sensor 1,2,3 in zone A and sensor 5,6,7 in zone B) and two \textit{ScioSense} \textit{CCS801} units (\cref{fig:processing}c, sensor 4 in zone C and sensor 8 in zone D). The sensing elements are designed and tuned to react to various gases, such as carbon monoxide, nitrogen dioxide, ammonia and hydrogen, as well as to a variety of VOCs, such as ethanol, methane, propane, iso-butane and acetone.
Each sensing element is interfaced through a dedicated low-noise operational amplifier (\textit{ST TS924}) and driven by a 12-bit, \SI{1}{\kilo\hertz} digital-to-analog converter (DAC, \textit{TI DAC60004}). Sensor resistances and heater currents are read back simultaneously via dual 24-bit, \SI{1}{\kilo\hertz} analog-to-digital converters (ADC, \textit{TI ADS131M08}), ensuring phase-coherent sampling across all channels. Environmental conditions (pressure, temperature, humidity) are monitored at \SI{50}{\hertz} using a \textit{TE MS8607} digital sensor. 
All electrical components are laid out on a four-layer printed circuit board (PCB) designed to minimise analogue coupling and electromagnetic interference. Power is delivered through a single \SI{5}{\volt} USB rail and managed to keep total consumption between \SI{1.2}{\watt} and \SI{1.5}{\watt}. Sensor heater control combines a feed-forward thermal model, calibrated against manufacturer specifications, with a proportional feedback loop to maintain each MOx hotplate at its setpoint. This per-sensor closed-loop arrangement compensates for variations in thermal mass, airflow and device ageing, ensuring stable baselines and rapid settling times. 
The microcontroller timestamps and streams raw sensor voltages, heater currents and environmental readouts into structured binary logs. A Python toolkit ingests these logs and enables signal post-processing. % baseline correction, feature extraction (e.g.\ windowed resistance differentials, frequency-domain transforms) and odour classification. 
The system has been shown to yield sub-\SI{100}{\milli\second} identification of brief odour pulses and supports temporal pattern decoding up to tens of hertz, thus matching or exceeding the temporal capabilities of mice on similar tasks \cite{dennler_high-speed_2024}.

\subsection*{Experimental protocol}
Different environmental flow conditions and odour source configurations were measured. 
We tested three flow velocities (\SI{10}{\centi \meter/\second}, \SI{15}{\centi \meter/\second}, and \SI{20}{\centi \meter/\second}), at three e-nose locations (Source centred (\texttt{C}, zone C\&D aligned \SI{27.1}{\centi\metre} from the source), \SI{5}{\centi\metre} upstream (\texttt{U}), and \SI{5}{\centi\metre} right (\texttt{R}), see \cref{fig:ex1}b). Every experimental condition was repeated twice. For each configuration, a background image (no tracer) and a laser flat-field image (homogeneous tracer concentration) %(see example in \cref{fig:processing}c) 
were acquired, which are later used to correct for additive background signals and spatial variations in illumination and detection efficiency.
The sensors on the e-nose were operated at a constant temperature of \SI{400}{\celsius}. Before the measurements, the sensors underwent a 24-hour pre-conditioning in ambient air, and a three-point heater calibration protocol to account for their state and ambient conditions. Before each measurement, the wind tunnel was flushed to remove any residual acetone vapour. The first PLIF image was timed to coincide with the start of the stimulus. Each acquisition lasted \SI{300}{\second}, from which the first \SI{25}{\second} were required for the odour source to stabilise. 
%
%The resulting 10 datasets are described in the following table:
%[...]

\subsection*{PLIF image processing}
The PLIF frames were initially saved as \textit{LaVision} “.imx” files, 
%first converted from the Andor “.sifx” format to \textit{LaVision} “.imx” using a MATLAB utility that employs the ATSIFIO reader and the readimx writer, 
with metadata (laser energy, timing, flow rates) preserved in accompanying header files. 
To obtain the 20 Hz time resolution, it was necessary to alternate laser head firing for each image, but due to the unique energy distributions of each laser head, we need to process the data from each separately for accurate concentration reconstruction. We therefore separated the image stacks from each laser head to carry out the following processing steps. 
In DaVis~10.2.1, each frame \(I_{\rm raw}(x,y,t)\) underwent background subtraction, whereby a time-averaged background image \(I_{\rm bg}(x,y)\) (laser on, no tracer) was subtracted to remove static reflections and dark current. 
Flat-field correction was then applied by dividing the result by the flat-field image \(F(x,y)\), acquired under constant tracer concentration, compensating for spatial variations in light sheet intensity. Prior to dividing, the background image was subtracted from the flat-field.
Next, to mitigate shot-to-shot laser energy fluctuations and inter-head intensity differences, each pixel was normalised by a calibration coefficient, \(a_c(t) \), derived from the instantaneous source intensity, \(I_{\rm src}(t)\), measured in a \SI{5}{\square\milli\metre} region at the injector exit. Together, this results in the source normalised concentration $C_{\rm norm}(x,y,t)$ as follows:
\[
C_{\rm norm}(x,y,t) = \frac{1}{a_c(t)} \frac{I_{\rm raw}(x,y,t) - I_{\rm bg}(x,y)}{F(x,y)-I_{\rm bg}(x,y)}
\]
%As illustrated in \cref{fig:processing}a, 
The temporal evolution of the source %\(I_{\rm src}(t)\) 
exhibits a ramp-up and stabilisation period of approximately 400 frames, and consequently the first 500 frames (\SI{25}{\second}) of each dataset were discarded. 
Residual optical distortions were subsequently removed via geometric unwarping using a mapping function based on images of a precisely machined calibration plate (the \textit{LaVision} dual-plane calibration plate 309–15), resulting in remapped image frames of size \num{1057} px$\times$\num{1058} px. Median spatial filtering with a 7$\times$7 px window was then applied to reduce noise while maintaining concentration gradients.
Finally, processed stacks from both laser heads were merged and exported, together with all timing and calibration parameters, into an HDF5 dataset for subsequent concentration field extraction and statistical analysis.

\subsection*{Spatial alignment}
To correlate the point-source e-nose measurements with the spatial PLIF data, the 2D PLIF concentration fields were converted into scalar time series by spatially averaging the source-normalised PLIF concentrations within predefined Regions of Interest (ROIs) corresponding to the physical locations and dimensions of the four MOx sensor housings. Each of the four sensor housings (labelled as zone \texttt{A}--\texttt{D}) was mapped to a rectangular ROI on the PLIF imaging plane (denoted as $\mathcal{P}_i$), as determined from a reference PLIF image (see \cref{fig:processing}a). The mean pixel intensity within this region was computed for every PLIF frame to obtain the sensor-specific concentration signal \(C'_i(t_{\mathrm{plif}}) = |\mathcal{P}_i|^{-1} \sum_{(x,y)\in\mathcal{P}_i} C(x,y,t_{\mathrm{plif}})\). This procedure transformed the PLIF image sequence into four synchronized time series, \(C'(t_{\mathrm{plif}}) = [C'_A(t_{\mathrm{plif}}), \ldots, C'_{D}(t_{\mathrm{plif}})]^{T}\), which were subsequently used to compare and align the PLIF-derived relative concentrations with the corresponding e-nose sensor responses.

\subsection*{Temporal alignment}
%To resolve the temporal mismatch between the high-frequency e-nose data and the slower PLIF acquisition rate, we implemented an upsampling procedure that aligns both signals on a shared time base. Synchronisation relied on a pulsed trigger signal recorded by the e-nose and embedded in the PLIF acquisition stream. Several interpolation schemes were assessed to estimate PLIF-derived concentrations at intermediate e-nose timestamps. Shape-preserving methods (PCHIP\cite{fritsch1980monotone} and Akima\cite{akima1970new}) were preferred for their ability to avoid non-physical artefacts such as overshoot and oscillation. A quantitative validation, based on splitting the PLIF time series into reconstruction and validation subsets, was used to benchmark each method via the Normalised Root Mean Square Error (NRMSE). Although linear interpolation yielded the lowest NRMSE, it was rejected due to its non-differentiable output. PCHIP was instead selected as the best compromise, offering low reconstruction error while retaining monotonicity and physical plausibility. The resulting PCHIP interpolant was applied to the full synchronised PLIF dataset to generate a high-resolution ground-truth concentration signal aligned with the e-nose measurements, enabling direct comparison of plume dynamics across modalities. An excerpt of the interpolated signal is shown in \cref{fig:processing}d.
To address the temporal mismatch between the high-frequency e-nose data and the lower-rate PLIF acquisition, both signals were aligned on a common time base using a pulsed trigger signal embedded in the PLIF acquisition stream and recorded by the e-nose. 
For enabling direct comparison of plume dynamics across the two modalities, PLIF-derived concentrations were interpolated and upsampled to the e-nose timestamps using a shape-preserving piecewise cubic Hermite interpolating polynomial (PCHIP) \cite{fritsch1980monotone}, selected to minimise non-physical artefacts such as overshoot or oscillation. The resulting interpolant was applied to the full synchronised PLIF dataset to generate a temporally aligned, high-resolution ground-truth concentration signal. An excerpt of the interpolated signal is shown in \cref{fig:processing}b.
For many applications, the native PLIF acquisition rate (\SI{20}{\hertz}) is sufficient to capture the dominant plume dynamics, particularly given the comparatively slow recovery times of MOx sensors. However, turbulent odour plumes exhibit fluctuations across a broad range of time scales, and recent work has shown that MOx sensors can encode structure close to millisecond resolution \cite{martinez_fast_2019, dennler_high-speed_2024, dennler_neuromorphic_2025}.
To preserve this potential high-frequency information and thus avoiding additional temporal aliasing introduced by resampling, we retain the temporally upsampled PLIF signal as the primary ground truth. The dataset explicitly distinguishes between original and interpolated measurements (see "Data Records"), and we provide scripts to facilitate user-defined temporal resampling for application-specific analyses (see "Usage Notes").

\section*{Data Records}
The dataset is publicly available through the \textit{ETH Research Collection} \cite{dennler2026_OdourSensingTurbulentPlumes}. For each of the 10 experiments, two data products are provided: (i) the full PLIF image sequence, and (ii) the combined e-nose and spatio-temporally aligned PLIF time series.
The full PLIF image sequences contain 5,500 processed frames per experiment together with all associated metadata (laser timing, flat-field information, calibration parameters). Owing to their heterogeneity and size, these data are stored in HDF5 format, with each file containing a hierarchical structure comprising the image stack, timing information, and acquisition metadata (see \cref{tab:h5_summary}).
%
% \begin{figure}[h!]
% \centering
% \begin{minipage}{0.9\linewidth}
% \small
% \begin{verbatim}
% /                               (HDF5 root)
% |-- DataSets                    (group)
% |   |-- R56_FinalData           (dataset, shape = 5500 x 1057 x 1058, float32)
% |   |-- background_laser3       (dataset)
% |   |-- background_laser4       (dataset)
% |   |-- flat-field_laser3        (dataset)
% |   `-- flat-field_laser4        (dataset)
% |
% |-- General_Info                (group)
% |
% `-- Mapping                     (group)
%     |-- odor source location    (dataset)
%     |-- scaling: pixels to mm   (dataset)
%     |-- x_grid                  (dataset)
%     `-- y_grid                  (dataset)
% \end{verbatim}
% \end{minipage}
% % \caption{Directory structure of the HDF5 file containing PLIF and auxiliary data.}
% \label{fig:h5_tree}
% \end{figure}
%
\begin{table}[hb!]
\centering
\begin{tabular}{ll}
\toprule
\textbf{Path} & \textbf{Description} \\
\midrule
\texttt{/DataSets/RXX\_FinalData} &
PLIF data cube (\(5500 \times 1056 \times 1046\), \texttt{float32}) \\
\texttt{/DataSets/background\_laser3 \& 4} &
Background image for laser 3 \& laser 4\\
% \texttt{/DataSets/background\_laser4} &
% Background image for laser 4 \\
\texttt{/DataSets/flat-field\_laser3 \& 4} &
Flat-field correction image for laser 3 \& laser 4\\
% \texttt{/DataSets/flat-field\_laser4} &
% Flat-field correction image for laser 4 \\
\midrule
\texttt{/General\_Info} &
Acquisition and experiment metadata \\
\midrule
\texttt{/Mapping/odor source location} &
Odour source reference position \\
\texttt{/Mapping/scaling: pixels to mm} &
Pixel-to-millimetre conversion factors \\
\texttt{/Mapping/x\_grid} &
X-coordinate grid (mm) \\
\texttt{/Mapping/y\_grid} &
Y-coordinate grid (mm) \\
\bottomrule
\end{tabular}
\caption{Summary of the groups and datasets contained in the HDF5 files.}
\label{tab:h5_summary}
\end{table}

\begin{table}[ht!]
\centering
\begin{tabular}{cccccccccc}
\toprule
\textbf{Timestamp} &
\textbf{Time (s)} &
\textbf{$R_{\text{gas},1}$} &
\textbf{$R_{\text{gas},2-8}$} &
\textbf{$T_{\text{heat},1}$} &
\textbf{$T_{\text{heat},2-8}$} &
\textbf{PLIF} &
\textbf{PLIF$_{A}$} &
\textbf{PLIF$_{C-D}$} &
% \textbf{$T$, $RH$, $p$} \\
\textbf{Env.} \\
\midrule
0 days 01:12:00.345447 &
0.000 &
11768.3 &
\ldots &
400.13 &
\ldots &
True &
0.0309&
... &
... \\
0 days 01:12:00.346448 &
0.001 &
11782.4 &
\ldots &
399.99 &
\ldots &
False &
0.0310 &
... &
... \\
\midrule
\multicolumn{10}{c}{\ldots (remaining dataset not shown)} \\
\bottomrule
\end{tabular}
\caption{Condensed representation of the multimodal dataset. Sensor and heater channels 2--8, PLIF zone \texttt{B}--\texttt{D} and the environmental variables (temperature, relative humidity, and pressure) are collapsed for readability.}
\label{tab:condensed_dataset}
\end{table}

The combined e-nose and PLIF data are provided as comma-separated value (CSV) files to enable efficient loading in common analysis environments. Each file contains synchronised e-nose resistance measurements (in Ohm), %~\si{\ohm})
heater temperatures (in~\si{\celsius}), a binary indicator marking whether a PLIF image was acquired ("True") or if the PLIF entries result from time-interpolation ("False"), the corresponding PLIF-derived concentration estimates for zone \texttt{A}--\texttt{D}, and environmental sensor readings (temperature in~\si{\celsius}, relative humidity in~\si{\percent}, and pressure in~\si{\milli\bar}) (see \cref{tab:condensed_dataset}). %The column structure is organised as follows:
%

%(This section should be used to explain what the dataset contains, including the repository where it is stored, an overview of the data files and their formats, and any folder structure. Each external dataset should be cited using our data citation format. Please do not include extensive summary statistics, which should be limited to less than half a page, with 1-2 tables or figures, if required at all. Note the general expectation is that, if readers wish to scrut+
%inise your dataset's contents, they will download and analyse it for themselves (meaning that if the figure may be generated from the data file we generally recommend it does not need to be summarised or depicted in the text). )

\begin{figure*}[hb!]
    \centering
    \includegraphics[width=0.955\textwidth]{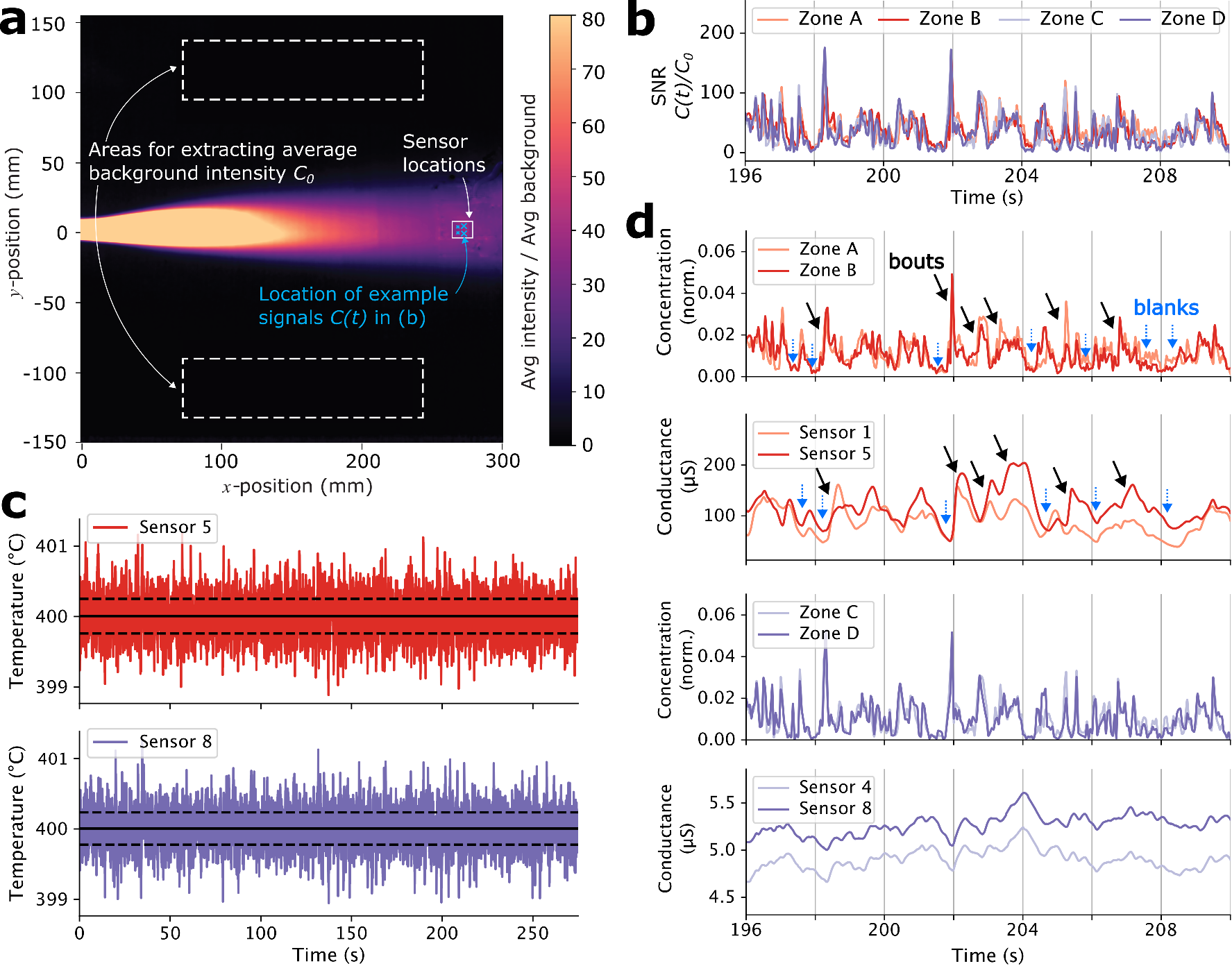}
    \caption{
    Technical validation. 
    \textbf{a} PLIF data, time-averaged signal-to-noise ratio (SNR). 
    \textbf{b} PLIF data, instantaneous SNR at different sensor locations. 
    \textbf{c} MOx hotplate temperature during experiment `\texttt{r56}', black solid and dotted lines correspond to mean and standard deviation, respectively.
    \textbf{d} Spatiotemporal alignment of PLIF concentration and sensor conductance measurements. Arrows indicate prominent turbulence features, illustrating the agreement between the modalities. 
    }
    \label{fig:validation}
\end{figure*}

\newpage
\section*{Technical Validation}
\subsection*{PLIF signal-to-noise evaluation}
The quality of PLIF data depends largely on the ability to induce and image high levels of acetone fluorescence (i.e., signal) relative to the background levels of light (i.e., noise). To this end, we use high-energy lasers with an output of approximately 85 mJ per pulse to maximize signal, while darkening the room and employing anti-reflective surfaces and optical bandwidth filtering to minimize noise. The processing procedure, as described previously, also includes steps to isolate the signal from background as well as account for potential artefacts such as light sheet variation and camera distortions. 
Computing the signal-to-noise ratio (SNR) of the data allows us to quantify the relative strength of the concentration measurements and provide technical validation for the dataset. To do so, we average the measurements of an area well outside the plume to obtain a value for the noise level, here $C_0=3*10^{-4}$. Then, we divide the average signal $C$ in each location by the noise level to obtain an average SNR. As shown in \cref{fig:validation}a, the average SNR over the sensor locations is greater than 30, providing confidence in the measured concentration. Additionally, this SNR is a conservative representation of the data quality, since the time-average of the fluctuating concentration measurements is generally lower than instantaneous concentration values $C(t)$. This can be seen in the sample time series shown in \cref{fig:validation}b, 
in which the instantaneous SNR regularly exceeds 100, indicating that plume fluctuations are well resolved above the noise floor.

\subsection*{MOx operating temperature stability}
The resistance of MOx sensors depends not only on the presence of gases, but also on the operating temperature. Air flow may directly affect the surface temperature of the sensors, which in turn may lead to signal degradation or artefacts when measuring gases. However, the e-nose used is equipped with fast temperature read- and set operations, and employs a dedicated closed-loop control system to maintain the sensor operating temperature constant. \cref{fig:validation}c displays the heater temperature across one full experiment (here Sensor 4 and 8 of experiment `\texttt{r56}'). For any given sensor and experiment, the temperature mean was ${\mu}_T=\SI{400.0}{\celsius}$ with a standard deviation of ${\sigma}_T\leq\SI{0.4}{\celsius}$.

\subsection*{MOx conductance values}
Because the experimental protocol relies on a deliberately strong acetone source, partial saturation of some sensors is unavoidable; several elements in the array exhibit high intrinsic sensitivity to acetone.  \cref{tab:gas_response_metrics} reports a dynamic-range retention score for each sensor across the different experimental runs, yielding the percentage of samples in an experiment that were non-saturated. A value of 100\% indicates that the sensor remained fully within its dynamic range throughout the measurement, and 0\% indicates that it was saturated for the entire duration. This metric effectively captures the usable operating range of each sensor under high-load conditions. The results show that Sensors 1, 4, 5, and 8 consistently maintain full dynamic range in all experiments and therefore provide robust signals for downstream analysis. By contrast, Sensors 2, 3, 6, and 7 display varying degrees of saturation, although it is noteworthy that in two runs (experiment \texttt{r70} \& \texttt{r71}, location \texttt{R}) none of these sensors reached saturation (likely due to their reduced exposure at a source-offset position). For subsequent modelling and interpretation, it is therefore advisable to prioritise the non-saturating sensors, as they offer stable performance and reliable quantitative information under the acetone concentrations encountered here.
%
% \begin{table}[ht]
% \centering
% \begin{tabular}{lcccccccc}
% \toprule
% {Experiment} & Sensor 1 & Sensor 2 & Sensor 3 & Sensor 4 & Sensor 5 & Sensor 6 & Sensor 7 & Sensor 8 \\
% \midrule
% 56 & 100.0\% & 98.1\% & 0.0\% & 100.0\% & 100.0\% & 14.2\% & 0.0\% & 100.0\% \\
% 57 & 100.0\% & 70.9\% & 0.0\% & 100.0\% & 100.0\% & 0.0\%  & 0.0\% & 100.0\% \\
% 58 & 100.0\% & 43.8\% & 0.0\% & 100.0\% & 100.0\% & 0.1\%  & 0.0\% & 100.0\% \\
% 59 & 100.0\% & 98.8\% & 0.0\% & 100.0\% & 100.0\% & 8.9\%  & 0.0\% & 100.0\% \\
% 60 & 100.0\% & 77.3\% & 0.0\% & 100.0\% & 100.0\% & 0.1\%  & 0.0\% & 100.0\% \\
% 61 & 100.0\% & 41.6\% & 0.0\% & 100.0\% & 100.0\% & 0.0\%  & 0.0\% & 100.0\% \\
% 70 & 100.0\% & 100.0\% & 100.0\% & 100.0\% & 100.0\% & 100.0\% & 100.0\% & 100.0\% \\
% \bottomrule
% \end{tabular}
% \caption{Dynamic-range retention score, expressed in percent.}
% \label{tab:sensor_saturation}
% \end{table}
%
\subsection*{Spatio-temporal alignment}
The spatio-temporal alignment has been performed via synchronising the data to shared trigger pulses, which were initiated by the PLIF acquisition software. Since the stimulus was not controlled externally, there are no sharp odour steps or pulses that can be used to validate the alignment. However, the turbulent nature of odour plumes produces distinct features\cite{celani_odor_2014}, which can be characterized as periods of time above some detectable threshold, known as `bouts', and periods of little or no signal, known as `blanks'. While the e-nose data features physical low-pass filtering characteristics, the response is fast enough such that these features can be recognised. In fact, \cref{fig:validation}d displays temporally and spatially aligned PLIF concentration and e-nose conductance data, where various `bouts' and `blanks' coincide in both modalities. The synchrony of these feature events is evident, where---as expected---the sensor conductance signal is smoother and lags slightly when compared to the PLIF concentration measurement. 
\begin{table}[ht!]
    \centering
\begin{tabular}{l l l r cccccccc}
        \toprule
        Exp.\ ID & Location & Velocity & \multicolumn{8}{c}{Dynamic-range retention score (\%)} \\
        \cmidrule(lr){4-11}
        & & (cm\,s$^{-1}$) 
        & $R_{\mathrm{gas},1}$ & $R_{\mathrm{gas},2}$ & $R_{\mathrm{gas},3}$ & $R_{\mathrm{gas},4}$ 
        & $R_{\mathrm{gas},5}$ & $R_{\mathrm{gas},6}$ & $R_{\mathrm{gas},7}$ & $R_{\mathrm{gas},8}$ \\
        \midrule
        r56 & C & 10 & 100.0 & 98.1 & 0.0 & 100.0 & 100.0 & 14.2 & 0.0 & 100.0 \\
        r57 & C & 15 & 100.0 & 70.9 & 0.0 & 100.0 & 100.0 & 0.0  & 0.0 & 100.0 \\
        r58 & C & 20 & 100.0 & 43.8 & 0.0 & 100.0 & 100.0 & 0.1  & 0.0 & 100.0 \\
        r59 & C & 10 & 100.0 & 98.8 & 0.0 & 100.0 & 100.0 & 8.9  & 0.0 & 100.0 \\
        r60 & C & 15 & 100.0 & 77.3 & 0.0 & 100.0 & 100.0 & 0.1  & 0.0 & 100.0 \\
        r61 & C & 20 & 100.0 & 41.6 & 0.0 & 100.0 & 100.0 & 0.0  & 0.0 & 100.0 \\
        % \addlinespace
        r70 & R & 10 & 100.0 & 100.0 & 100.0 & 100.0 & 100.0 & 100.0 & 100.0 & 100.0 \\
        r71 & R & 10 & 100.0 & 100.0 & 100.0 & 100.0 & 100.0 & 100.0 & 100.0 & 100.0 \\
        % \addlinespace
        r75 & U & 10 & 100.0 & 92.0 & 0.0 & 100.0 & 100.0 & 2.1 & 0.0 & 100.0 \\
        r76 & U & 10 & 100.0 & 85.0 & 0.0 & 100.0 & 100.0 & 2.2 & 0.0 & 100.0 \\
        \bottomrule
    \end{tabular}
    \caption{Dynamic-range retention score, defined as the percentage of samples in an experiment that were not saturated, for different sensors and experiments. Locations refer to the areas described in \cref{fig:ex1}b; velocity refers to the wind tunnel air flow.}
    \label{tab:gas_response_metrics}
\end{table}

\begin{figure*}[b!]
    \centering
    \includegraphics[width=0.99\textwidth]{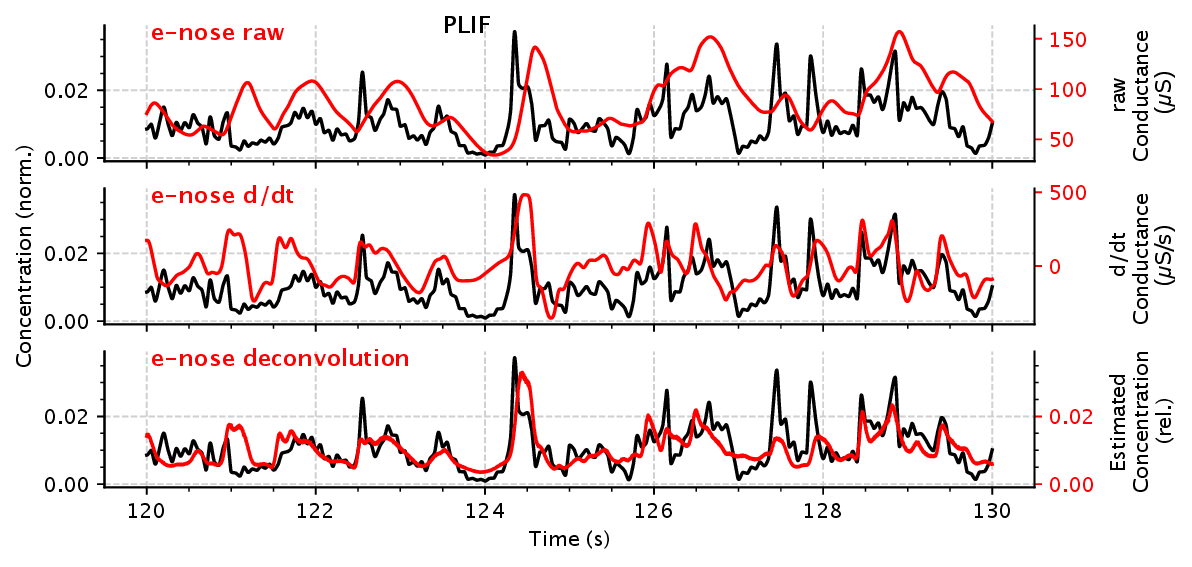}
    \caption{
    Usage notes: Time-series comparison between the PLIF-derived relative concentration (black) and co-located e-nose measurements (red). Top: raw MOx conductance, middle: temporal derivative of the conductance + phase alignment, bottom: output of a supervised deconvolution approach  \cite{martinez_fast_2019} + phase alignment.
    }
    \label{fig:usage}
\end{figure*}

\section*{Usage Notes}
\subsection*{Data handling}
The combined e-nose and spatio-temporally aligned PLIF time series can be accessed directly from CSV files (each <100 MB), enabling straightforward integration into standard analysis workflows.
For the raw high-resolution PLIF image sequences, efficient data handling is recommended when working with long recordings. To facilitate this, we provide a set of Python scripts and Jupyter notebooks for loading, inspecting, and processing the dataset. These examples illustrate typical workflows, including navigation of the directory structure, access to metadata, and selective loading of PLIF data via chunking, frame selection, and temporal or spatial subsampling; enabling efficient processing on standard hardware. Furthermore, for the combined dataset, different subsampling strategies are provided, which allow for either the original \SI{20}{\hertz} data, or the interpolated data (as described in the Methods under "Temporal alignment").
The notebooks can be executed without modification to reproduce the figures shown in this manuscript, or adapted for user-specific analyses. For integration into larger pipelines, the provided scripts include entry points for exporting PLIF fields to NumPy arrays, xarray datasets, or other common scientific data structures. All files use open, platform-independent formats to support reuse across a wide range of computational environments.
%
% %
% Due to the high-resolution PLIF image sequences, efficient handling is recommended when working with long recordings. The notebooks illustrate strategies such as chunked loading, frame selection, and downsampling for exploratory work, as well as methods for mapping image timestamps to the corresponding e-nose samples. For users who wish to integrate the dataset into larger pipelines, the scripts provide entry points for exporting PLIF fields to NumPy arrays, xarray datasets, or other common scientific data formats. All files use open, platform-independent formats to support reuse in a variety of computational environments.
% %
% Further, the example scripts include routines for visualisation and quality control, such as plotting concentration fields, inspecting laser-sheet intensity corrections, and checking the synchronisation between optical and sensor measurements. %These tools are intended as practical guidelines for dataset interaction and can be extended or replaced with alternative workflows as required.

\subsection*{Estimating gas concentration changes from the e-nose conductance}
A common workflow when using this dataset may be to evaluate algorithms that estimate gas concentration changes from e-nose signals. Because the e-nose measurements and PLIF fields are co-located and aligned through validated synchronisation procedures, users can directly load the corresponding time series from a single dataset and apply their preferred estimation method without requiring additional preprocessing. We provide an accompanying Jupyter notebook that illustrates this workflow, including data access, temporal alignment, and example processing steps.
\cref{fig:usage} shows an example in which the relative concentration obtained from the PLIF data is compared with e-nose signals after several standard transformations. These are intended as technical demonstrations of how users might structure their own analyses.
For instance, overlaying the raw conductance illustrates the baseline sensor dynamics, while deriving the first temporal derivative or applying a published supervised deconvolution method \cite{martinez_fast_2019} demonstrates how different signal-processing choices affect agreement with the optical measurements. Other approaches, such as system modelling \cite{monroy_overcoming_2012}, band-pass filtering \cite{schmuker_exploiting_2016}, blind deconvolution \cite{martinez_fast_2019}, Kalman Filtering \cite{drix_resolving_2021}, or reinforcement learning strategies \cite{husnain_gas_2024}, may be tested by modifying the estimation block in the notebook.
%

% One of the most obvious use cases of this dataset is to benchmark different algorithms that estimate the actual gas concentration from the e-nose output. Due to the co-localisation and the validated spatio-temporal data alignment, this is as simple as loading two columns from the same dataset, and feeding these to the algorithm of choice. 
% %
% We provide an additional Jupyter notebook demonstrating the use. 
% %
% \cref{fig:validation} displays the relative concentration measured via PLIF, and overlays the e-nose after undergoing different processing steps. While a substantial mismatch can be seen using the naive approach, i.e. overlaying the raw e-nose conductance with the PLIF data, significant improvements are evident when either using the first derivative of the conductance, or the output of a prominent supervised deconvolution approach \cite{martinez_fast_2019}. 
% %
% Further algorithms could be tested in a similar fashion, i.e. by replacing the estimation step in the provided notebook. 
%('Usage Notes' is an optional section that can be used to provide information that may assist other researchers who reuse your data. Most commonly these are additional technical notes about how to access or process the data. Please do not use this section to write a conclusions section, general selling points, worked cases studies, or similar, as we do not publish these.)

\section*{Data and Code Availability}
All datasets generated and described in this work are publicly available through the \textit{ETH Research Collection} \cite{dennler2026_OdourSensingTurbulentPlumes}. %Each recording includes PLIF image sequences, e-nose time series, metadata, calibration files, and aligned data products in HDF5 format.
All code used for data processing, temporal and spatial alignment, and example analyses is provided in a public Git repository at \url{https://github.com/molecularsensing/enose-plif-datadescriptor}. %The repository includes Python scripts and Jupyter notebooks for loading, inspecting, and reproducing the workflows presented in this Data Descriptor.

\section*{Competing Interests}
The authors declare that the research was conducted in the absence of any commercial or financial relationships that could be construed as a potential conflict of interest.

\section*{Acknowledgements}
N.D. and A.T.G. acknowledge funding from the Swiss National Science Foundation (Grant SNF \#10001714), the ETHZ Postdoctoral Fellowship (Grant \#25-1 FEL-031), and the Swiss State Secretary for Education, Research and Innovation (MB22.00041, ERC-STG-21 "HealthSense"). 
%
%M.S.\ acknowledges funding from the NSF/MRC NeuroNex Odor2Action programme (NSF \#2014217, MRC \#MR/T046759/1).
%
M.S., A.T., J.P.C., L.L., and E.S. acknowledge funding by the National Science Foundation (USA), under the NSF/CIHR/DFG/FRQ/UKRI-MRC Next Generation Networks for Neuroscience Program (award number 2014217 to JPC and MRC \#MR/T046759/1) and the NSF GRFP (Grant number DGE2040434 to EAS).
\bibliography{sn-bibliography}% common bib file


\end{document}